\documentclass[conference,onecolumn,10pt]{IEEEtran}

\author{\IEEEauthorblockN{Giacomo Bergami}
\IEEEauthorblockA{Dept. of Computer Science\\
Free University of Bozen-Bolzano\\
Bozen, Italy\\
Email: bergamigiacomo@gmail.com}}

\title{A Logical Model for joining Property Graphs}

\usepackage{graphicx}
\newcommand{\ignore}[1]{}
\usepackage{booktabs}
\usepackage{amsmath,amsthm}
\newtheorem{lemma}{Lemma}
\newtheorem{example}{Example}
\newtheorem{definition}{Definition}
\usepackage{amssymb}
\usepackage{ifsym}
\usepackage{paralist}
\DeclareMathOperator{\dom}{dom}

\usepackage{pifont}
\usepackage{braket}\usepackage{paralist}
\usepackage[inline]{enumitem}
\newlist{myalist}{enumerate*}{1}
\setlist[myalist]{label=\textbf{(\arabic*)}}
\newlist{mylist}{enumerate*}{1}
\setlist[mylist]{label=\textbf{(\roman*)}}
\newlist{alphalist}{enumerate*}{1}
\setlist[alphalist]{label=\textbf{(\alph*)}}

\usepackage{xcolor}
\usepackage{listings}
\definecolor{eclipseBlue}{RGB}{42,0.0,255}
\definecolor{eclipseGreen}{RGB}{63,127,95}
\definecolor{eclipsePurple}{RGB}{127,0,85}
\lstdefinelanguage{sparql}{
	morecomment=[l][]{\#},
	morestring=[b][]\",
	morekeywords={BIND,URI,CONCAT,SELECT,CONSTRUCT,DESCRIBE,ASK,WHERE,FROM,NAMED,PREFIX,BASE,OPTIONAL,FILTER,GRAPH,LIMIT,OFFSET,SERVICE,UNION,EXISTS,NOT,BINDINGS,MINUS,a},
	sensitive=true
}
\lstdefinelanguage{cypher}{
	morekeywords={MATCH,RETURN,WHERE,DISTINCT,WITH,CREATE,COUNT,AS,UNION,ALL,is,null,NOT,AND,OR},
	sensitive=true,
	morecomment=[l]{//}, 
}

\begin{document}
\maketitle

\begin{abstract}
The present paper upgrades the logical model required to exploit materialized views over property graphs as intended in the seminal paper ``A Join Operator for Property Graphs''. Furthermore, we provide some computational complexity proofs strengthening the contribution of a forthcoming graph equi-join algorithm proposed in a recently accepted paper. 
\end{abstract}
\bigskip

The present technical reports addresses some of the reviewers' questions for \cite{BergamiBM21} that, despite being interesting, don't fit in the original paper.

\section{{{Preliminaries}}}
{The \textbf{image} of a function $f\colon A\to B$ is the set $\Im f$ of all output values it may produce. Formally, $\Im f=\Set{f(x)|x\in A}$. The \textbf{power set} $\wp(S)$ of any set $S$ is the set of all subsets of $S$, including the empty set $\emptyset$ and $S$ itself. Formally, $\wp(S)=\Set{S'|S'\subseteq S}$}

{Given a set of attributes $A$ and a set of values $\nu$, a \textbf{tuple} is a finite function $t\colon A\mapsto \nu$, where each attribute $a\in A$ is either associated to a value in $\nu$ or is undefined ($a\notin\textup{dom}(t)$). An empty tuple $\varepsilon$ has an empty domain. We denote $\mathcal{T}$ as the set of all the possible tuples.}

{A \textbf{multiset} is a pair $(S,\mu)$, denoted as $S_\mu$, where $S$ is the \textit{underlying set} of the multiset, formed from its distinct elements, and $\mu\colon S\mapsto \mathbb{N}_{\geq 1}$ associates each element $s\in S$ to the number of its occurrences in the multiset $\mu(s)$. We can represent a multiset as an \textit{indexed set} $\overline{S_\mu}\subseteq S\times \Im \mu$ where each element $s_i\in S_\mu$ represents the $i$-th replica of $s\in S$:
	${\overline{S_\mu} = \Set{s_i\in S\times \Im \mu|0<i\leq \mu(s)}}$. Relational tables might be then defined as indexed sets of tuples.}

{ Given $\mathcal{T}_{\mu}$ a indexed set of tuples, {a set of attributes $A$, and a set of labels $\Sigma$}, a \textbf{property graph} is a tuple $G=(V_{\mu'},E_{\mu''},\lambda,\ell)$
	where \begin{alphalist}
		\item $V_{\mu'}\subseteq \mathcal{T}_{\mu}$ is an indexed set of nodes,
		\item $E_{\mu''}\subseteq \mathcal{T}_{\mu}$ is an indexed set of edges with $\overline{V_{\mu'}}\cap \overline{E_{\mu''}}=\emptyset$,\\
		
		\item $\lambda\colon{\mathcal{T}_\mu\to \mathcal{T}_\mu\times \mathcal{T}_\mu}$ is a function assigning node pairs to edges,
		\item $\ell\colon{\mathcal{T}_\mu}\to \wp(\Sigma)$ is a function assigning a set of labels to nodes {and edges} 
\end{alphalist} }

\section{Logical Model}
{\textit{Property graphs} are directed, labelled and attributed multigraphs. With reference to Figure 1 from \cite{BergamiBM21}, a collection of \textit{labels}  is associated to every vertex and edge (e.g., \texttt{\{Paper\}} or \texttt{\{Cites\}}). Vertices and edges may have arbitrary named attributes (\textit{properties}) in the form of key-value pairs (e.g., \texttt{\textsc{Title}=\textit{Graphs}} or \texttt{\textsc{1Author}=\textit{Alice}}).}  We decide to embed the relational model on property graphs, so that the standard operators' properties from the relational model can be inherited: {so, property-value associations of vertices (and edges) in indexed sets $V_{\mu'}$ (and $E_{\mu''}$) can be represented by tuples $t$ and a number $i$, so that $t_i$ identifies the $i$-th istance of $t$.}

\begin{example}\label{ex:onJoinDataModel}
	Using the same figure as a reference, two vertices \ding{172} and \ding{176} come from two distinct property graphs, respectively Researcher and Citation. Such vertices have the following property-value associations:
	\[\text{\ding{172}}(\textsc{Name})=\textit{Alice}\qquad \text{\ding{176}}(\textsc{1Author})=\textit{Alice}\qquad \text{\ding{176}}(\textsc{Title})=\textit{Graphs}\]
	Given $\ell$ the function associating each vertex to its label set, we also have $\ell(\text{\ding{172}})=\{\textup{User}\}$ and $\ell(\text{\ding{176}})=\{\textup{Paper}\}$.
	Edge \textit{i} connects vertex \ding{172} to \ding{173} ($\lambda(\textit{i})=(\text{\ding{172}} ,\text{\ding{173}})$), while \textit{v} connects \ding{176}
	to \ding{178} ($\lambda(\textit{v})=(\text{\ding{176}} ,\text{\ding{178}})$). Given $\ell$ the edge labelling function, 
	such edges store only the following labels:
	\[\ell(\textit{i})=\Set{\textup{Follows}},\quad\ell(\textit{v})=\Set{\textup{Cites}}\]
\end{example}

Considering that the standard property graph model used in some graph databases such as Neo4J cannot represent a graph database as a collection of graphs, we can represent 
each graph within the collection as a distinct connected component within a single graph (database). This can be formalised as follows:

\begin{definition}[Graph Database]
	{A \textbf{graph database} is a single property graph, where each $i$-th graph operand is represented as one single connected component of vertices in $\mathcal{V}(i)$ and edges in $\mathcal{E}(i)$, where $\mathcal{V}$ (and $\mathcal{E}$) maps each $i$-th component to a subset of $V_{\mu'}$ (and $E_{\mu''}$). }
\end{definition}

{The combination of objects lies at the basis of the graph join operations, and therefore influences the definition of the logical data model. When two vertices (or edges) are combined into one, the resulting vertex (or edge) must contain all the information that was initially stored in the two original vertices (or edges), thus including the property-value associations and the labels. We then need to define the following function}:
\begin{definition}[Combination]
	A \textbf{combination operator}  $\oplus:A\cdot A\mapsto A$ is a  function between two 
	operands of type $A$ returning an element of the same type, $A$.  $\oplus$ is defined for the following $A$-s:
	\begin{itemize}
		\item \textbf{sets}: it performs the set union:\qquad
		$S\oplus S'\overset{def}{=}S \cup S'$
		\item {\textbf{integers from finite sets $M,N$}: given $i\in M$ and $j\in N$, it returns the dovetail number, associating each pair of
			integers with an unique integer:
			$i\oplus j\overset{def}{=}(\max\{\max M,\max N\} + 1) + \sum_{k=0}^{i+j}k+\min\{i,j\}$}
		\item \textbf{functions}: given  functions $f:A\mapsto B$ and $g:C\mapsto D$, $f\oplus g$ is the overriding of $f$ by $g$ returning $g(x)$ if $x\in \dom(g)$, and $f(x)$ if $x\in \dom(f)$.
		
		\item \textbf{pairs}: given two pairs $(u,v)$ and $(u',v')$, then
		the pair combination is defined as the pairwise {{combination}} of each component, that is $(u,v)\oplus (u',v')\overset{def}{=}(u\oplus u',v\oplus v')$. 
		{Given that members of indexed sets  $s_i,t_j\in \overline{S_\mu}$ are pairs where integers $i$ and $j$ belong to a finite set $\Im \mu$, we can write $s_i\oplus t_j\overset{def}{=}(s\oplus t)_{i\oplus j}$.\footnote{In this case, please note that $i\oplus j=(\max \{\max\Im \mu,\max \Im\mu'\} + 1)+\sum_{k=0}^{i+j}+\min\{i,j\}=(\max\{\max_x\mu(x) ,\max_y\mu'(y)\}+1)+\sum_{k=0}^{i+j}+\min\{i,j\}$}}
	\end{itemize}
\end{definition}
{Given that both vertices and edges are defined as tuples (i.e., finite functions), we might update the property-value associations via a relational $\theta$-Join between two indexed sets of vertices or edges. The relational $\theta$-Join is then defined as follows:}

\begin{definition}[$\theta$-Join]\label{def:thetajoin}
	Given two {indexed sets of tuples} $R$ and $S$, the \textbf{$\theta$-join} $R\bowtie_\theta S$ in \cite{atzeniIT}  is defined as {a natural join followed by a selection operation under the $\theta$ predicate\footnote{The formal definition of the operation is provided in \S\ref{app:reljoins}.}}:
	If $\theta$ is the always true predicate, $\theta$ {{can}} be omitted.
\end{definition}

{Let us now focus on updating the label information. Given the usual definition of a property graph,  the vertex labelling function $\ell$ might contain the label information for vertices \ding{172} and \ding{176}, but it might not contain the labelling information for vertex $\text{\ding{172}}\oplus\text{\ding{176}}$ resulting from the join operation, as such vertex is generated at run-time. For this reason, we need to extend both the labelling functions  and $\lambda$ at run time. We can therefore define the following function:}

\begin{definition}[Run-time extension]\label{def:rte}{
		Given a function $f\colon A\mapsto B$ and a  combination operator $\oplus$ defined over both $A$ and $B$, a \textbf{run-time extension} for a finite function $f$ is a function $F_f$ that returns $f(z)$ if $f$ is defined over $z$, returns $F_f(x)\oplus F_f(y)$ if and only if $z\equiv x\oplus y$ and $f$ is defined over both $x$ and $y$, and otherwise it is undefined\footnote{Formally, $F_f(z)=\begin{cases}
			f(z) & z\in\textup{dom}(f)\\
			F_f(x)\oplus F_f(y) & z\notin\textup{dom}(f),z\equiv x\oplus y,x,y\in\textup{dom}(F_f)
			\end{cases}$}. This function {requires that $\oplus$ is lazily evaluated} so  that $x\oplus y$ can be still decomposed into $x$ and $y$ before recursively applying $F$.}
	
	{When $A_{\mu'}$ is an indexed set of tuples ($\overline{A_{\mu'}}\subset \overline{\mathcal{T}_\mu}$),  we can verify that $\dom(F_f)$ can be written as the following non-finite enumerable domain: $A_{\mu'}\cup \bigcup_{n\in\mathbb{N}_{\geq 1}}\bowtie_{i\leq n}A_{\mu'}$. We can shorthand this set as $\wp^\oplus(A)$.}
\end{definition}

\begin{example}
	By continuing Example \ref{ex:onJoinDataModel}, suppose now that the edge $\textit{i}\oplus \textit{v}$ comes from
	a graph join where edges from Researcher are joined to the ones in Citation in a resulting graph,
	where  vertices $\text{\ding{172}}\oplus \text{\ding{176}}$
	and $\text{\ding{173}}\oplus \text{\ding{178}}$  also appear. So:
	\[(\text{\ding{172}}\oplus\text{\ding{176}})(\textsc{Name})=\textit{Alice}\;\;\; (\text{\ding{172}}\oplus\text{\ding{176}})(\textsc{1Author})=\textit{Alice}\;\;\; (\text{\ding{172}}\oplus\text{\ding{176}})(\textsc{Title})=\textit{Graph}\]
	By {exploiting the definition of the run-time extension}, the labels are combined with set unions:
	\[\begin{split}
	F_{\ell}(\textit{i}\oplus \textit{v})=\ell(\textit{i})\oplus\ell(\textit{v})=\{\textup{Follows}\}\oplus\{\textup{Cites}\}=\{\textup{Follows},\textup{Cites}\}\\
	\end{split}\]
	Also, the joined edge $\textit{i}\oplus \textit{v}$ will point to the joined vertices as requested, i.e. $F_\lambda(\textit{i}\oplus\textit{v})=\lambda(\textit{i})\oplus\lambda(\textit{v})=(\text{\ding{172}},\text{\ding{173}})\oplus (\text{\ding{176}},\text{\ding{178}})=(\text{\ding{172}}\oplus\text{\ding{176}},\text{\ding{173}}\oplus \text{\ding{178}})$.
	As we will see in the incoming properties section, these requirements allow the graph conjunctive and disjunctive joins to be both commutative and associative. 
\end{example}

{We can now extend the whole property graph at run-time (i.e., a materialized view) as follows:}

\begin{definition}[Run-time extendible Property Graph]\label{def:rtepg}
	{Given a property graph $G=(V_\mu,E_{\mu'},\lambda,\ell)$, a \textbf{run-time extendible property graph}  is a property graph $G_F=(V_\mu,E_{\mu'},F_{\lambda},F_{\ell})$ having the same vertices, edges, attributes, and labels of $G$, but where the vertex and edge labelling function is $F_{\ell}$  and where the edge-vertex association is $F_{\lambda}$.}
\end{definition}

Since all of the relevant information is stored in the graph database, we represent 
the graph as the set of the minimum information required for the join operation.

\begin{definition}[Graph]
	The $i$-th \textbf{graph} of a graph da\-tabase {represented as a run-time extendible property graph} $\mathcal{D}$ 
	is a tuple
	$G_i=(\mathcal{V}(i),\mathcal{E}(i))$, where $\mathcal{V}(i)$ is a multiset of vertices and 
	$\mathcal{E}(i)$ is a multiset of edges. 
\end{definition}

\section{Graph $\theta$-Joins}\label{gj}
Given that id-joins briefly presented in the introduction are a specific case of the $\theta$-joins, we  only formalise the latter.  $G_a\bowtie_\theta^{\textup{\textbf{es}}}G_b$
expresses the join of graph $G_a$ with $G_b$, where:
\begin{enumerate}
	\item we first use a relational $\theta$-join among the vertices,
	and then \item we combine the edges using an appropriate user-determined
	edge semantics, \textbf{es}\index{semantics!edge (graph join)}.
\end{enumerate}
This  modularity is similar to
the operators previously described in graph theory literature, where instead of a join between vertices they have a
cross product, and different semantics are expressed as different
graph products. Please note that this operation definition on the logical model must not necessarily describe how the actual graph join algorithm works, but must only provide a suitable representation for analysing the graph joins' desired properties. The general graph $\theta$-join for the logical model is defined as follows:

\begin{definition}[Graph $\theta$-Join]\label{def:join}
	\label{def:graphjoin} \index{product, $\otimes_\theta$!for property graphs!join}\index{$\bowtie$!property graphs|see {product $\otimes_\theta$}}
	Given two graphs $G_a=(V,E)$ and $G_b=(V',E')$ within the same graph database $D$, a \textbf{graph
		$\theta$-join} is defined as follows:
	\begin{equation*}
	G_a\bowtie_\theta^{\textbf{\textup{es}}} G_b=(V\bowtie_\theta V',{\textup{\textbf{es}}(E,E')})
	\end{equation*}
	where $\theta$ is a binary predicate over the vertices and $\bowtie_\theta$ is the $\theta$-join (Definition \ref{def:thetajoin}) among the vertices, and
	${\textup{\textbf{es}}}$ maps the edges $(E,E')$ to the subset of all the possible edges linking the vertices in $V\bowtie_\theta V'$.
\end{definition}
Given that graph join
returns a property graph like the graphs in input, property graphs are
closed under the graph join operator via the definition of $\oplus$ for the multiset
$\theta$-join. 
The result of the join between two graphs, a Researcher Graph and References, produces the same
set of vertices regardless of the edge semantics of choice. On the other hand, edges among the resulting vertices change according to the edge semantics. In the first one  we  combine edges appearing in both graphs
and link vertices that appear combined in the resulting graph. We have a \textbf{Conjunctive Join}, that in graph theory is known as \textit{Kronecker graph product}.
In this case  $E_{\textup{\textbf{es}}}$ is defined with the ``$\wedge$'' \textbf{es} semantics as an edge join
$E_{\wedge}=E\bowtie_{\Theta_\wedge}E'$, where the ${\Theta_\wedge}$ predicate is the following:
\begin{equation}
\label{eq:conj}
\Theta_\wedge(e_h,e_k')=(e_h\in E\wedge e_k'\in E') \wedge F_\lambda(e_h\oplus e_k')\in (V\bowtie_\theta V')^2
\end{equation}

We can also define a \textbf{disjunctive} semantics, having ``$\vee$'' as \textbf{es}.
In this case, we want edges appearing either in the first or in the second operand.
This means that two vertices, $u_h\oplus u'_h$ and $v_k\oplus v'_k$, {{could}} have a resulting
edge
$e_i\oplus \varepsilon'_j$
even if only $\lambda(e_i)=(u_h,v_k)$ appears in the first operand\footnote{The statement addressing the edges in the second operand that do not bond with the ones in the other graph is expressed as follows: $\forall e'\in E'. \neg\Theta_\wedge(e_i,e')$}
and $\varepsilon'$ is a ``fresh'' empty edge $\lambda(\varepsilon'_j)=(u'_h,v'_k)$ not
appearing in $G_b$ such that $F_\lambda(e_a\oplus \varepsilon'_b)=(u_h\oplus u'_h,v_k\oplus v'_k)$.
Consequently the disjunctive join can be defined as follows:
\begin{equation}
\label{eq:disj}
\begin{split}
E_{\vee}&=E{\bowtie}_{\Theta_\wedge}E'\\
&\quad\scriptstyle  \cup\{(e_i\oplus\varepsilon_j)|\varepsilon_j,\,\left(\forall e'\in E'. \neg\Theta_\wedge(e_i,e')\right), \ell(\varepsilon_j)=\emptyset,\;\lambda(e_i)\oplus (v,v')\in \mathbf{V}_{\bowtie}   \}\\
&\quad\scriptstyle \cup\{(\varepsilon_i\oplus e_j)|\varepsilon_i,\,\left(\forall e\in E. \neg\Theta_\wedge(e,e_j)\right), \ell(\varepsilon_i)=\emptyset,\;(u,u')\oplus \lambda(e_j)\in \mathbf{V}_{\bowtie}  \}\quad \\
\end{split}
\end{equation}
where $\mathbf{V}_{\bowtie}:=(V\bowtie_\theta V')^2$


\subsection{Graph $\theta$-Joins: Properties}\label{gjprop}
The following subsection motivates the specific choice of these two edge semantics for joining graphs: we want to show that graph joins are scalable with respect to the generalisation of such joins to multiple graphs. First, this requires, as previously discussed, that the proposed graph $\theta$-join is closed under composition.  Second, graph joins (for each given semantics) must be commutative and associative: these graph join properties allow a future scalable implementation of multi-way graph joins. In fact, we could implement such graph multi-joins by chaining binary graph joins, and starting the computation from the smallest graph operand up to the greatest one, such that we potentially reduce the number of possible join comparisons.

\begin{lemma}[Join Commutativity]
	Given two graphs $G$ and $G'$ from the same graph database $\mathcal{D}$ and a symmetric binary predicate
	$\theta$, we have $G\bowtie_\theta^\wedge G' \equiv G'\bowtie_{\theta^{-1}}^\wedge G$ for the conjunctive
	semantics and $G\bowtie_\theta^\vee G' \equiv G'\bowtie_{\theta^{-1}}^\vee G$ to the disjunctive one.
\end{lemma}
\begin{proof}
	Similarly to the relational algebra case where this definition is required\footnote{See Appendix \ref{app:reljoins} for a formal proof of the relational algebra commutativity over indexed sets.}, we {have}  $\theta^{-1}$ as the {symmetric}  predicate of $\theta$, such that $\theta^{-1}(b,a)\Leftrightarrow \theta(a,b)$. For the conjunctive semantics, we have that $G\bowtie_\theta^\wedge G'$ is:
	\[(V\bowtie_\theta V',E\bowtie_{\Theta_\wedge}E')\]
	Since we have that $V\bowtie_\theta V'=V'\bowtie_{\theta^{-1}} V$ and
	$E\bowtie_{\Theta_\wedge}E'=E'\bowtie_{\Theta_\wedge}E$ {for the relational algebra}, then we have that
	the graph join is equivalent to $G'\bowtie_{\theta^{-1}}^\wedge G$.
	
	This is proved because the relational join between vertices and edges is  a commutative operator
	\cite{rolleke94equivalencesof}, and predicate $\Theta_\wedge$  is symmetric
	when either $E$ and $E'$ or $E'$ and $E$ are joined. A similar proof could be carried out for
	the disjunctive semantics by expanding the definition of the disjunctive edge semantics.
	The resulting equation will be true, because the relational join and the $\oplus$ operator with the null tuple are symmetric operators.
\end{proof}

The following corollary strengthens the previous result: it shows that join commutativity implies
having two resulting graphs, where both vertices and edges have the same labels and the same edges
link the same vertices.

\begin{lemma}[Commutativity for $\lambda$ {and} $\ell$]\label{coroll:Comm}
	For each vertex $v_i\oplus v_j'$ from the vertex set $V\bowtie_\theta V'$ from
	$G\bowtie_\theta^\wedge G'$ and the corresponding equivalent vertex $v_j'\oplus v_i$ in $V'\bowtie_\theta V$ from
	$G'\bowtie_\theta^\wedge G$ ($v_i\oplus v_j'=v_j'\oplus v_i$), we have that both vertices have the same label set.
	
	For the conjunctive semantics, for each edge $e_h\oplus e_k'$ from the edge set $E\bowtie_{\Theta_\wedge} E'$ from
	$G\bowtie_\theta^\wedge G'$ and the corresponding equivalent edge $e_k'\oplus e_h$ in $E'\bowtie_{\Theta_\wedge} E$ from
	$G'\bowtie_\theta^\wedge G$ ($e_h\oplus e_k'=e_k'\oplus e_h$), we have that both edges have the same label set and link
	the same equivalent vertices. This statement also applies for the disjunctive semantics.
\end{lemma}
\begin{proof}
	This corollary is proved by the linearity of run-time extensions. Regarding the vertex labelling, we have that the labelling provided by the result of the two commutated joins is the same by the commutativity of the set union operator:
	\[\begin{split}
	F_{\ell}(u_i\oplus u_j')&=\ell(u_i)\oplus \ell(u_j')=\ell(u_i)\cup\ell(u_j')=\\
	&=\ell(u_j')\cup\ell(u_i)=\ell(u_j')\oplus\ell(u_i)=\\
	&=F_{\ell}(u_j'\oplus u_i)
	\end{split}
	\]
	
	Regarding the conjunctive semantics, the proof of $\ell(e_h\oplus e_k')=\ell(e_k'\oplus e_h)$ is similar by using the set union's commutativity. We  prove that equivalent edges link equivalent vertices:
	\[\begin{split}
	F_{\lambda_{E_\wedge}}(e_h\oplus e_k')&=F_{\lambda_E}(e_h)\oplus F_{\lambda_{E'}}(e_k')=(u_i,v_j)\oplus (u_l',v_m')=\\
	&=(u_i\oplus u_l',v_j\oplus v_m')=(u_l'\oplus u_i,v_m'\oplus v_j)
	\end{split}  \]
	\[F_{\lambda_{E_\wedge}}(e_k'\oplus e_h)=F_{\lambda_{E'}}(e_k')\oplus F_{\lambda_{E'}}(e_k')=(u_l'\oplus u_i,v_m'\oplus v_j)  \]
	The proofs for the disjunctive semantics are the same.
\end{proof}

Since the relational algebra $\theta$-join operator satisfies associativity \cite{rolleke94equivalencesof}, we could carry out a similar
proof for join associativity:

\begin{lemma}[Join Associativity]
	Given three graphs $G$, $G'$ and $G''$ from the same graph database $\mathcal{D}$ and a symmetric binary predicate
	$\theta$,  we have $G\bowtie_{\theta_1\wedge\theta_\alpha}^\wedge (G'\bowtie_{\theta_2}^\wedge G'') = (G\bowtie_{\theta_1}^\wedge G')\bowtie_{\theta_\alpha\wedge \theta_2}^\wedge G''$ for the conjunctive
	semantics and $G\bowtie_{\theta_1\wedge\theta_\alpha}^\vee (G'\bowtie_{\theta_2}^\vee G'') = (G\bowtie_{\theta_1}^\vee G')\bowtie_{\theta_\alpha\wedge\theta_2}^\vee G''$ for the disjunctive one.
\end{lemma}
\begin{proof}
	Since we have that the usual $\theta$-relational joins are associative as outlined by the following equivalence:
	\[(A\bowtie_{\theta_1}B)\bowtie_{\theta_\alpha\wedge \theta_2}C =A\bowtie_{\theta_1\wedge\theta_\alpha}(B\bowtie_{\theta_2}C)\]
	then, we have that the relational $\theta$-joins among the edges are associative too, as well as the theta joins among the edges. Hereby, the join between the graphs is associative. 
\end{proof}

Similarly to the graph join's commutativity, we can strengthen the result for the join associativity with the following corollary:

\begin{lemma}[Associativity for $\lambda$ {and} $\ell$]\label{coroll:Assoc}
	For each vertex $v_i\oplus (v_j' \oplus v_k'')$ from the vertex set $V\bowtie_{\theta_1\wedge\theta_\alpha}^\wedge (V'\bowtie_{\theta_2}^\wedge V'')$ from
	$G\bowtie_{\theta_1\wedge\theta_\alpha}^\wedge (G'\bowtie_{\theta_2}^\wedge G'')$ and the corresponding equivalent vertex $(v_i\oplus v_j')\oplus v_k''$ in $V'\bowtie_\theta V$ from
	$(G\bowtie_{\theta_1}^\wedge G')\bowtie_{\theta_\alpha\wedge \theta_2}^\wedge G''$ ($v_i\oplus (v_j'\oplus v_k'')=(v_i\oplus v_j')\oplus v_k''$), we have that both vertices have the same label set.
	
	For the conjunctive semantics, for each edge $e_h\oplus (e_k'\oplus e_t'')$ from the edge set $E\bowtie_{\Theta_\wedge} (E'\bowtie_{\Theta_\wedge}E'')$ from
	$G\bowtie_{\theta_1\wedge\theta_\alpha}^\wedge (G'\bowtie_{\theta_2}^\wedge G'')$ and the corresponding equivalent edge $(e_h\oplus e_k')\oplus e_t''$  from
	$ (G\bowtie_{\theta_1}^\wedge G')\bowtie_{\theta_\alpha\wedge \theta_2}^\wedge G''$, we have that both edges have the same label set and link
	the same equivalent vertices. This statement also applies to the disjunctive semantics.
\end{lemma}
\begin{proof}
	This corollary is proved by the linearity of run-time extensions. Regarding the vertex labelling, we have that the labelling provided by the result of the two commutated joins is the same by the associativity of the set union operator:
	\[\begin{split}
	F_{\ell}(u_i\oplus (u_j'\oplus u_k''))&=\ell(u_i)\oplus F_{\ell}(u_j'\oplus u_k'')\\
	&=F_{\ell}(u_i\oplus u_j')\oplus\ell(u_k'')\\
	&=F_{\ell}((u_i\oplus u_j')\oplus u_k'')
	\end{split}
	\]
	
	Regarding the conjunctive semantics, the proof of $e_h\oplus (e_k'\oplus e_t'')=(e_h\oplus e_k')\oplus e_t''$ is similar, by using the set union's associativity. We  prove that equivalent edges link equivalent vertices:
	\[\begin{split}
	F_{\lambda_{E_\wedge}}((e_h\oplus e_k')\oplus e_t'')&=(u_i\oplus u_l',v_j\oplus v_m')\oplus(u_n'',u_p'')=\\
	&=(u_i\oplus u_l'\oplus u_n'',v_j\oplus v_m'\oplus u_p'')\\ & = F_{\lambda_{E_\wedge}}(e_h\oplus (e_k'\oplus e_t''))
	\end{split}  \]
	The proofs for the disjunctive semantics are the same.
\end{proof}

\section{Graph Equi-Join Algorithm}
\subsection{Proofs} This section provides the proofs associated to the novel equi-join algorithm proposed in our forthcoming paper \cite{BergamiBM21}.

\begin{lemma}
	Given two graph operands $G_a$ and $G_b$ and a $\theta$ binary operator, the conjunctive algorithm runs in time $T_\wedge(G_a,G_b)\in O(|\mathcal{V}(a)||\mathcal{V}(b)||\mathcal{E}(a)||\mathcal{E}(b)|)$ in the worst case scenario, and is $T_\wedge(G_a,G_b)\in O(|\mathcal{E}(a)\cup \mathcal{E}(b)|+|\mathcal V(a)|\log|\mathcal V(a)|+|\mathcal V(b)|\log |\mathcal V(b)|)$ in the best case scenario.
\end{lemma}
\begin{proof}
	We first describe the computational cost of each phase of the provided algorithm.
	
	\textbf{Loading}: for each operand $G_x$, the $map_x$ construction  takes at most time $\sum_{j=0}^{|\mathcal V(x)|}\log(j)$, where $|\mathcal V(x)|$ is the multi-set vertex size. Such time complexity       is bounded by $|\mathcal V(x)|\leq \sum_{j=0}^{|\mathcal V(x)|}\log(j)<|\mathcal V(x)|\log|\mathcal V(i)|$ for $|\mathcal V(x)|> 5$. The outgoing edges sorting costs $|v.\textit{out}_x|+|v.\textit{out}_x|\cdot \log |v.\textit{out}_x|$ for each vertex $v$.
	
	\textbf{Indexing}: given $k_x$ the size of \textsc{Keys}($map_x$), the serialisation phase takes $3k_x+|\mathcal{V}(x)|+|\mathcal{E}(x)|$ time, where $2k_x$ is the red-black tree map visit cost, $k_x$ is the $map_x$ serialisation cost as \textit{HashOffset} and $|\mathcal{V}(x)|+|\mathcal{E}(x)|$ is the time to serialise the graph as \textit{VertexVals[]}.
	
	\textbf{ConjunctiveJoin}: Given the graph operands $G_a$ and $G_b$, this last operation takes time $\max\{k_a,k_b\}+\sum_{h\in \textit{BI}}\left( b_a^h\cdot b_b^h+out_a^h\cdot out_b^h\right)$ where $b_x^h$ is the size of the $h$-th bucket for the $x$-th operand,
	while $out_x^h$ is the outgoing vertices' size for all the vertices within the $h$
	bucket for the $x$-th operand.
	\medskip
	
	In the worst case scenario, each operand $G_x$ is a complete graph ($out_x^h=\mathcal{E}(x)$) and the hashing function is a constant: each vertex is mapped into a single bucket. Therefore, $k_a=k_b=1$ and $b_x^h=|\mathcal{V}(x)|$: the resulting computational complexity is $|\mathcal{V}(a)||\mathcal{V}(b)||\mathcal{E}(a)||\mathcal{E}(b)|$, in compliance with the relational query plan showed in Figure 4 in \cite{BergamiBM21}. In this scenario the computation cost is dominated by \textsc{ConjunctiveJoin}.
	
	In the best case scenario, for each operand $G_x$ each bucket is ``small'' ($b_x^h\ll |\mathcal{V}(x)|$) and the number of outgoing edges for each vertex is constant ($out_x^h\ll |\mathcal{E}(x)|$). 
	Therefore, the loading phase has a quasi-linear cost of $O(|\mathcal V(x)|\log|\mathcal V(i)|)$. As a result,  vertices dominate in the loading operation, which is quasi-linear with respect to the size of the vertices; the edges also sensibly influence the indexing operation for dense graphs.
\end{proof}

\begin{lemma}
	With respect to the time complexity, the best (worst) case scenario of the disjunctive semantics is asymptotically equivalent to the conjunctive semantics in its best (worst) case scenario, i.e.  {${\lim\sup}_{(|G_a|,|G_b|)\to (+\infty, +\infty)}\frac{T_\wedge(|G_a|,|G_b|)}{T_\vee(G_a,G_b)}=1$}, under the same algorithmic conditions {, where $|G_x|$ is a shorthand for $|\mathcal{V}(x)|+|\mathcal{E}(x)|$}.
\end{lemma}
\begin{proof}
	As we can see from the Graph EquiJoin Algorithm, the disjunctive semantics provides additional edges that are not considered in the conjunctive one. Therefore, the disjunctive semantics can at least have the same computational complexity of the conjunctive semantics. We need to determine if and when such conditions can be met. If for each matching vertex  $(v\oplus v')$ such that $\theta(u,v')$ we have $\mathtt{Keys}(v.map)=\mathtt{Keys}(v'.map)$, then the \textsc{Join} function is not going to visit more edges than the ones already visited for the conjunctive semantics; in addition to that, we need to determine if and when \textsc{Disjunction} is never going to be called: we can see that this condition is met for $E_L=E_R=\emptyset$. Given that these conditions could be met, the disjunctive semantics has its best case scenario when all the aforementioned conditions hold and, as a result, it has the same computational complexity of the conjunctive semantics. In this case, the limit condition is trivially met. 
	
	In the worst case scenario we have that, before calling \textsc{Disjunction}, $|E_L|\in O(out_a^h)$ ($|E_R|\in O(out_b^h)$) for each $h\in \textit{BI}$ and that $\mathtt{Keys}(v.map)\neq \mathtt{Keys}(v'.map)$. Given that we implemented $E_L$ and $E_R$ as unordered sets, insertion and removal elements from such sets are $O(1)$. This implies that the computational complexity of \textsc{Disjunction} is $b_a^h|E_R|+b_b^h|E_L|$; in the worst case scenario, which conditions are the same as those in the previous lemma, this computational complexity becomes $O(|\mathcal{V}(a)||\mathcal{E}(b)|+|\mathcal{V}(b)||\mathcal{E}(a)|)$. Last, we need to add this contribution to the conjunctive semantics' computational complexity: given that  $|\mathcal{V}(a)||\mathcal{E}(b)|+|\mathcal{V}(b)||\mathcal{E}(a)|\in O(T_\wedge(G_a, G_b))$ under the worst case scenario for the conjunctive semantics, we have that the limit condition is met even in this occasion. {The goal limit reduces to:
	} \[
	\lim {\sup}
	{_{(|G_a|,|G_b|)\to (\infty, \infty)}\frac{T_\wedge(G_a,G_b)}{T_\vee(G_a,G_b)}=}
	{\lim\sup}
	{_{(|G_a|,|G_b|)\to (+\infty, +\infty)}}   \frac{1}{1+\frac{|\mathcal{V}(a)||\mathcal{E}(b)|+|\mathcal{V}(b)||\mathcal{E}(a)|}{|\mathcal{V}(a)||\mathcal{V}(b)||\mathcal{E}(a)||\mathcal{E}(b)|}}=1
	\]
\end{proof}
The experiment section of the associated paper shows that, despite $T_\vee(G_a,G_b)\in O(T_\wedge(G_a,G_b))$ for the previous lemma, the contribution of $T_\vee(G_a,G_b)-T_\wedge(G_a,G_b)$ is going to be relevant in Big Data scenarios. 

\subsection{Detailed Experimental Results}
\begin{table*}[!h]
	\centering
	\caption{Loading time over the enriched Friendster and Kroneker Datasets. 1H=3.60$\cdot 10^3$ s}
	\resizebox{\textwidth}{!}{%
		\begin{tabular}{l|ll|ll|ll|ll}
			\toprule
			$|V_1|=$ & \multicolumn{2}{c}{\textbf{Proposed Loading}}& \multicolumn{2}{|c|}{\textbf{PostgreSQL+SQL}}&  \multicolumn{2}{|c|}{\textbf{Virtuoso+SPARQL}}& \multicolumn{2}{|c|}{\textbf{Neo4j+Cypher}}\\
			$|V_2|$ & Friendster (s)          & Kroneker (s)          & Friendster (s)          & Kroneker (s)& Friendster  (s)          &  Kroneker (s)& Friendster (s)          & Kroneker (s)         \\
			\midrule
			$10^1$ & \textbf{5.20} $\cdot 10^{-4}$ & \textbf{4.88} $\cdot 10^{-3}$ & 3.38 $\cdot 10^{-2}$          & 3.50 $\cdot 10^{-2}$    & 2.49 $\cdot 10^{-2}$ & 4.09 $\cdot 10^{-2}$ & 6.12 $\cdot 10^{-1}$ & 7.10 $\cdot 10^{-1}$ \\
			$10^2$ & \textbf{1.58} $\cdot 10^{-3}$ & \textbf{8.84} $\cdot 10^{-3}$ & 3.62 $\cdot 10^{-2}$          & 4.34 $\cdot 10^{-2}$   & 1.36 $\cdot 10^{-1}$ & 1.68 $\cdot 10^{-1}$ & 6.79 $\cdot 10^{-1}$ & 5.95 $\cdot 10^{-1}$\\
			$10^3$ & \textbf{1.22} $\cdot 10^{-2}$ & \textbf{3.33} $\cdot 10^{-2}$ & 4.73 $\cdot 10^{-2}$          & 4.70 $\cdot 10^{-2}$   & 1.21 $\cdot 10^0$ & 1.29 $\cdot 10^0$ & 9.36 $\cdot 10^{-1}$ & 8.81 $\cdot 10^{-1}$\\
			$10^4$ & \textbf{1.12} $\cdot 10^{-1}$         & \textbf{3.01} $\cdot 10^{-1}$ & {1.27}$\cdot 10^{-1}$ & 8.57 $\cdot 10^{-1}$  & 1.15 $\cdot 10^1$ & 1.09 $\cdot 10^1$& 2.72 $\cdot 10^{0}$  & 6.63 $\cdot 10^0$\\
			$10^5$ & 1.22 $\cdot 10^{0}$          & 2.95 $\cdot 10^{0}$  & \textbf{5.13}$\cdot 10^{-1}$ & \textbf{9.35} $\cdot 10^{-2}$      & 1.07 $\cdot 10^2$ & 1.37 $\cdot 10^3$ & 1.60 $\cdot 10^{1}$  &  1.77 $\cdot 10^1$ \\
			$10^6$ & {3.20} $\cdot 10^2$            & 2.95 $\cdot 10^{1}$    & \textbf{1.39}$\cdot 10^1$    & \textbf{4.97} $\cdot 10^{0}$    & {\color{blue}$>1$H} & {\color{blue}$>1$H} & 1.64 $\cdot 10^{2}$  &  1.77 $\cdot 10^2$\\
			
			$10^7$ & 6.93 $\cdot 10^2$            & 3.02 $\cdot 10^{2}$    & \textbf{5.33}$\cdot 10^1$    & \textbf{6.65} $\cdot 10^{1}$  & {\color{blue}$>1$H} &  {\color{blue}$>1$H} & 4.86 $\cdot 10^2$    & {\color{red}OOM1}\\
			
			$10^8$ & {7.69} $\cdot 10^2$          & 9.93 $\cdot 10^2$    & \textbf{2.20}$\cdot 10^{1}$   & \textbf{1.86} $\cdot 10^{2}$    & {\color{blue}$>1$H} & {\color{blue}$>1$H} & 4.62 $\cdot 10^2$    & {\color{red}OOM1}\\
			\bottomrule           
	\end{tabular}}
	\label{tab:1a}
\end{table*} 
\begin{table}[!h]
	\centering
	\caption{\textbf{Graph Equi-Joins} over the enriched Friendster Dataset: (Indexing and) Join Time. 1H=3.60$\cdot 10^3$ s. In {\color{blue}blue} we remark the benchmarks that exceeded the one hour threshold, while in {\color{red}red} we remark whether the algorithm failed to compute due to either primary ({\color{red}OOM1}) or secondary  ({\color{red}OOM2}) out of memory error (both set at 50GB of free space).}
	\resizebox{\textwidth}{!}{%
		\begin{tabular}{l|ll|ll|ll|ll}
			\toprule
			$|V_1|=$ & \multicolumn{2}{c}{\textbf{Proposed Join Algorithm}}& \multicolumn{2}{|c|}{\textbf{PostgreSQL+SQL}} & \multicolumn{2}{|c|}{\textbf{Virtuoso+SPARQL}}& 
			\multicolumn{2}{|c|}{\textbf{Neo4j+Cypher}}\\
			$|V_2|$ & Disj. (s)          & Conj. (s)          & Disj. (s)          & Conj. (s)& Disj.  (s)          & Conj. (s)	& Disj. (s)          & Conj. (s)         \\
			\midrule
			$10^1$ & \textbf{1.43} $\cdot 10^{-4}$ & \textbf{1.26}$\cdot 10^{-3}$ & \ignore{$\quad$}1.63$\:\;\cdot\; 10 {}^{-2}$ & 6.32 $\cdot 10^{-3}$   &   2.46 $\cdot 10^{0}$ & 3.18 $\cdot 10^{-1}$ & {\color{blue}$>$1H} & 5.37 $\cdot 10^{-2}$ \\
			$10^2$ & \textbf{1.97} $\cdot 10^{-3}$ & \textbf{1.72}$\cdot 10^{-3}$ & \ignore{$\quad$}1.67$\:\;\cdot\; 10 {}^{-2}$           & 1.63 $\cdot 10^{-2}$  &  3.55 $\cdot 10^{0}$ & 3.59 $\cdot 10^{-1}$ & {\color{red}OOM1}  & 1.04 $\cdot 10^0$\\
			$10^3$ &  \textbf{1.42} $\cdot 10^{-2}$ & \textbf{1.35}$\cdot 10^{-2}$ &  \ignore{$\quad$}6.46$\:\;\cdot\; 10 {}^{-2}$ & 2.40 $\cdot 10^{-2}$&   1.82 $\cdot 10^{1}$ & 1.57 $\cdot 10^0$ & {\color{red}OOM1}  & 2.50 $\cdot 10^0$\\
			$10^4$ &  \textbf{1.09} $\cdot 10^{-1}$ & \textbf{1.03}$\cdot 10^{-1}$ & \ignore{$\quad$}6.48$\:\;\cdot\; 10 {}^0$  & 8.57 $\cdot 10^{-1}$   
			&  1.84 $\cdot 10^{2}$ & 2.06 $\cdot 10^1$& {\color{red}OOM1}  & 2.57 $\cdot 10^1$\\
			$10^5$ &  \textbf{1.11} $\cdot 10^{0}$   & \textbf{1.12}$\cdot 10^{0}$  &  \ignore{\color{red}$>$9.52$^{*}\cdot{10^2}$}{\color{red}OOM2} & 7.11 $\cdot 10^{1}$    
			&   {\color{blue}$>1$H} & 1.37 $\cdot 10^3$ & {\color{red}OOM1}  & {\color{blue}$>1$H} \\
			$10^6$ &  \textbf{2.01} $\cdot 10^{1}$  & \textbf{1.11}$\cdot 10^1$    & \ignore{\color{red}$>$1.52$^{*}\cdot{10^3}$}{\color{red}OOM2} & 5.27 $\cdot 10^2$     
			&  {\color{blue}$>1$H}  & {\color{blue}$>1$H} &  {\color{red}OOM1}  & {\color{blue}$>1$H}\\
			$10^7$ &  \textbf{2.62} $\cdot 10^{2}$  & \textbf{5.08}$\cdot 10^1$    & \ignore{\color{red}$>$3.06$^{*}\cdot {10^{3}}$}{\color{red}OOM2}  & 3.30 $\cdot 10^{3}$  
			&    {\color{blue}$>1$H} & {\color{blue}$>1$H} &  {\color{red}OOM1}   & {\color{blue}$>1$H}\\
			$10^8$ &  \textbf{1.55} $\cdot 10^{3}$  & \textbf{1.34}$\cdot 10^2$    & \ignore{\color{red}$>$3.06$^{*}\cdot {10^{3}}$}{\color{red}OOM2}   & 3.38 $\cdot {10^{3}}$  
			&  {\color{blue}$>1$H}  & {\color{blue}$>1$H} &  {\color{red}OOM1}  & {\color{blue}$>1$H}\\
			\bottomrule           
	\end{tabular}}
	\label{tab:1b}
\end{table} \begin{table}[!h]
	\centering
	\caption{\textbf{Graph Equi-Joins}: Over the enriched Kronecker Graph Dataset: (Indexing and) Join Time. 1H=3.60$\cdot 10^3$ s: dashes remark when the join experiment cannot be executed because the dataset loading time went out of memory.}
	\resizebox{\textwidth}{!}{%
		\begin{tabular}{l|ll|ll|ll|ll|}
			\toprule
			$|V_1|=$ & \multicolumn{2}{c}{\textbf{Proposed GCEA Algorithm}}& \multicolumn{2}{|c|}{\textbf{PostgreSQL+SQL}} & \multicolumn{2}{|c|}{\textbf{Virtuoso+SPARQL}}& \multicolumn{2}{|c|}{\textbf{Neo4j+Cypher}}\\
			$|V_2|$ & Disj. (s)          & Conj. (s)          & Disj. (s)          & Conj. (s)        &  Disj. (s)          & Conj. (s)& Disj. (s)          & Conj. (s)         \\
			\midrule
			$10^1$ & \textbf{3.44} $\cdot 10^{-4} $ & \textbf{3.31} $\cdot 10^{-4} $& \ignore{$\quad$}6.84$\:\;\cdot\; 10 {}^{-3}$  &  5.74 $\cdot 10^{-3}$& 2.30 $\cdot 10^{0}$ & 1.84 $\cdot 10^{-1}$ & {\color{blue}$>1$H} & 5.76 $\cdot 10^{-1}$\\
			$10^2$ & \textbf{1.01} $\cdot 10^{-3}$ & \textbf{1.04} $\cdot 10^{-3}$ &  \ignore{$\quad$}8.63$\:\;\cdot\; 10 {}^{-3}$  & 1.59 $\cdot 10^{-2}$  & 2.31 $\cdot 10^{0}$ & 2.73 $\cdot 10^{-1}$ & {\color{red}OOM1}  & 1.32 $\cdot 10^0$\\
			$10^3$ & \textbf{8.07} $\cdot 10^{-3}$ & \textbf{7.17} $\cdot 10^{-3}$ &  \ignore{$\quad$}9.89$\:\;\cdot\; 10 {}^{-1}$ & 1.82 $\cdot 10^{-2}$ &  2.89 $\cdot 10^{0}$  & 9.47 $\cdot 10^{-1}$ &{\color{red}OOM1} & 2.55 $\cdot 10^0$ \\
			$10^4$ & \textbf{8.00} $\cdot 10^{-2}$& \textbf{8.13} $\cdot 10^{-2}$& \ignore{$\quad$}1.25$\:\;\cdot\; 10 {}^1$ & 6.61 $\cdot 10^{-1}$  &   2.47 $\cdot 10^{1}$   & 1.87 $\cdot 10^1$ &  {\color{red}OOM1}    & 3.20 $\cdot 10^1$\\
			$10^5$ & \textbf{9.19} $\cdot 10^{-1}$  & \textbf{9.32} $\cdot 10^{-1}$& \ignore{\color{red}$>$1.76$^{*}\cdot{10^3}$}{\color{red}OOM2} & 8.21 $\cdot 10^{1}$ &   1.77 $\cdot 10^3$   & 1.59 $\cdot 10^3$ &    {\color{red}OOM1} & {\color{blue}$>1$H}\\
			$10^6$ & \textbf{1.01} $\cdot 10^2$  & \textbf{9.95} $\cdot 10^{1}$& {\color{red}OOM2}   & 5.18 $\cdot 10^{2}$  &    {\color{blue}$>1$H}  & {\color{blue}$>1$H} &  {\color{red}OOM1}    & {\color{blue}$>1$H}\\
			$10^7$ & \textbf{5.21} $\cdot 10^2$ & \textbf{1.22} $\cdot 10^2$ &  {\color{red}OOM2} & {\color{blue}$>1$H}    &    {\color{blue}$>1$H}   & {\color{blue}$>1$H} &  {\color{red}OOM1}    & {\color{red}--} \\
			$10^8$ &   {\color{blue}$>1$H} & \textbf{8.96} $\cdot 10^2$&  {\color{red}OOM2}  & {\color{blue}$>1$H}   &    {\color{blue}$>1$H}   & {\color{blue}$>1$H} &  {\color{red}OOM1}    & {\color{red}--}\\
			\bottomrule           
	\end{tabular}}
	\label{tab:2}
\end{table} 

\newpage
\appendix

\subsection{{Relational joins over indexed sets are commutative}}\label{app:reljoins}

We can formally define the relational join over tuples' indexed sets as follows:
\[\overline{R_\mu}\bowtie_\theta\overline{S_{\mu'}}=\Set{r_i\oplus s_j|r_i\in \overline{R_\mu}, s_i\in \overline{S_{\mu'}}, \theta(r_i,s_i), \forall x\in \dom(r)\cap\dom(s). r(x)=s(x)}\]
Please remember that $\forall x\in \dom(r)\cap\dom(s). r(x)=s(x)$ always holds by vacuous truth when $\dom(r)\cap\dom(s)=\emptyset$.
\begin{lemma}\label{proof:OfCommut}
	Relational joins over indexed sets are commutative, i.e., $\overline{R_\mu}\bowtie_\theta\overline{S_{\mu'}}=\overline{R_\mu}\bowtie_{\theta^{-1}}\overline{S_{\mu'}}$, where $\theta^{-1}(a,b)\leftrightarrow \theta(b,a)$.
\end{lemma}
\begin{proof}
	The proof boils down to prove that, given $r_i\oplus s_j\in \overline{R_\mu}\bowtie_\theta\overline{S_{\mu'}}$ and $s_j\oplus r_i\in \overline{R_\mu}\bowtie_{\theta^{-1}}\overline{S_{\mu'}}$, we have $r_i\oplus s_j=s_j\oplus r_i$ where (H1) $\theta(r_i,s_i)\wedge \forall x\in \dom(r)\cap\dom(s). r(x)=s(x)$ and (H2) $\theta^{-1}(s_i,r_i)\wedge \forall x\in \dom(s)\cap\dom(r). s(x)=r(x)$ are both verified. We can see that conditions (H1) and (H2) are equivalent by $\theta^{-1}$ definition and for equivalence's reflexivity: we can prove (H1) and ignore (H2). 
	
	Given that such elements are pairs, this boils down to prove that both $i\oplus j=j\oplus i$ and $r\oplus s=s\oplus r$ holds when (H1) holds.
	
	Starting from the integers $i\oplus j=j\oplus i$ defined over the sets $\Im \mu$ and $\Im \mu'$, by expanding the definition of $\oplus$ we have that:
	\[\begin{split}
	i\oplus j &= \max\{\max \Im \mu,\max \Im \mu'\}+1+\sum_{k=0}^{i+j}k+\min\{i,j\}\\
	&= \max\{\max \Im \mu', \max \Im \mu\}+1+\sum_{k=0}^{i+j}k+\min\{j,i\}\\
	&= j\oplus i\\
	\end{split}\]
	
	For functions $r\oplus s=s\oplus r$, we can expand their definition as follows:
	\[\forall x\in\dom(r)\cup\dom(s).\; (r\oplus s)(x)=\begin{cases}
	s(x) & x\in\dom(s)\\
	r(x) & x\in\dom(r)\\
	\end{cases}\overset{?}{=}\begin{cases}
	r(x) & x\in\dom(r)\\
	s(x) & x\in\dom(s)\\
	\end{cases} = (s\oplus r)(x)\]
	We have the following cases:
	\begin{itemize}
		\item when both $r$ and $s$ are empty tuples, $\dom(r)=\dom(s)=\emptyset$, then we have that also $r\oplus s=s\oplus r$ will be empty tuples.
		\item when at least one tuple among $r$ and $s$ is non-empty but their domain intersection is empty, the above condition boils down to the following, which is equivalent:
		\[\begin{cases}
		s(x) & x\in\dom(s)\backslash\dom(r)\\
		r(x) & x\in\dom(r)\backslash\dom(s)\\
		\end{cases}{=}\begin{cases}
		r(x) & x\in\dom(r)\backslash\dom(s)\\
		s(x) & x\in\dom(s)\backslash\dom(r)\\
		\end{cases}\]
		\item when at least one tuple among $r$ and $s$ is non-empty as well as their domain intersection, the above condition boils down to the following, which is equivalent:
		\[\begin{cases}
		s(x) & x\in\dom(s)\backslash\dom(r)\\
		s(x) & x\in\dom(s)\cap\dom(r)\\
		r(x) & x\in\dom(r)\\
		\end{cases}\overset{?}{=}\begin{cases}
		r(x) & x\in\dom(r)\backslash\dom(s)\\
		r(x) & x\in\dom(r)\cap\dom(s)\\
		s(x) & x\in\dom(s)\\
		\end{cases}\]
		Given that $H1$ must hold because such tuples are returned as part of the join operation, then the left part of the is equivalent to the right one because we can apply $\forall x\in\dom(r)\cap\dom(s). r(x)=s(x)$ to the second condition.
	\end{itemize}
\end{proof}

Similar proofs can be done for the join asosciativity over indexed sets

\subsection{Graph Query Languages limitations' on Graph Joins}
The reason of comparing our graph join with multiple graph query languages is twofold: we want both to show that graph joins can be represented in different data representations (RDF and Property Graphs), and
to detail how our experiments in {\cite{BergamiBM21}} were performed.\\

At the time of the writing, graph query languages do not provide
	a specific keyword to operate the graph join between two graphs. Moreover, all the current graph
	query languages, except SPARQL, assume that the underlying GDBMS stores only one graph at a time, and hence
	binary graph join operations are not supported. As a consequence, an operator over one single graph operand must be implemented instead.
	
	Let us now suppose that we want to express a specific $\theta$-join, where the binary predicate $\theta$
	is fixed, into a (Property) Graph Query Language:
	in this case we have both \textit{(i)} to specify how a final merged vertex $v\oplus v''$ is obtained from each
	possible pair of vertices containing different possible attributes $A_1,A_2,\dots,A_n$ , and \textit{(ii)} to discard
	the pair of vertices that do not jointly satisfy  the
	$\theta$ predicate and the following \textit{join condition} (that varies upon the different vertices'
	attributes over the graph):
	\[(v\oplus v'')(A_1)=v\wedge  (v\oplus v'')(A_2)=v''\]
	
	\begin{figure}[!t]
		\begin{minipage}[t]{\textwidth}
\begin{lstlisting}[language=cypher,basicstyle=\ttfamily\scriptsize]
MATCH (src1)-[:r]->(dst1),
(src2)-[:r]->(dst2)
WHERE src1.Organization1=src2.Organization2 AND src1.Year1=src2.Year2 AND dst1.Organization1=dst2.Organization2 AND dst1.Year1=dst2.Year2 AND src1.graph='L' AND src2.graph='R' AND dst1.graph='L' AND dst2.graph='R'
CREATE p=(:U {Organization1:src1.Organization1, Organization2:src2.Organization2 , Year1:src1.Year1, Year2:src2.Year2 ,  MyGraphLabel:"U-"})-[:r]->(:U {Organization1:dst1.Organization1, Organization2:dst2.Organization2 , Year1:dst1.Year1, Year2:dst2.Year2, MyGraphLabel:"U-"}) return p
UNION ALL
MATCH (src1)-[:r]->(u), (src2)-[:r]->(v)
WHERE src1.Organization1=src2.Organization2 AND src1.Year1=src2.Year2 AND src1.graph='L' AND src2.graph='R' AND ((u.Organization1<>v.Organization2 OR u.Year1<>v.Year2))
CREATE p=(:U {Organization1:src1.Organization1, Organization2:src2.Organization2 , Year1:src1.Year1, Year2:src2.Year2 , MyGraphLabel:"U-"}) return p
UNION ALL
MATCH (src1)-[:r]->(u), (src2)
WHERE src1.Organization1=src2.Organization2 AND src1.Year1=src2.Year2 AND src1.graph='L' AND src2.graph='R' AND (NOT ((src2)-[:r]->()))
CREATE p=(:U {Organization1:src1.Organization1, Organization2:src2.Organization2 , Year1:src1.Year1, Year2:src2.Year2 , MyGraphLabel:"U-"}) return p
UNION ALL
MATCH (src1), (src2)-[:r]->(v)
WHERE src1.Organization1=src2.Organization2 AND src1.Year1=src2.Year2 AND src1.graph='L' AND src2.graph='R' AND (NOT ((src1)-[:r]->()))
CREATE p=(:U {Organization1:src1.Organization1, Organization2:src2.Organization2 , Year1:src1.Year1, Year2:src2.Year2 , MyGraphLabel:"U-"}) return p
UNION ALL
MATCH (src1), (src2)
WHERE src1.Organization1=src2.Organization2 AND src1.Year1=src2.Year2 AND src1.graph='L' AND src2.graph='R' AND (NOT ((src2)-[:r]->())) AND (NOT ((src1)-[:r]->()))
CREATE p=(:U {Organization1:src1.Organization1, Organization2:src2.Organization2 , Year1:src1.Year1, Year2:src2.Year2 , MyGraphLabel:"U-"}) return p
\end{lstlisting}
			\caption{Cypher implementation for the graph conjunctive equi-join operator. Please note that one of the limitations of such query language is that each vertex and edge is going to be visited and path-joined more than one time for each pattern where it appears.}
			\label{fig:CypherEquiJoin}
		\end{minipage}
	\end{figure}
\begin{figure}
		\begin{minipage}[!t]{\textwidth}
\begin{lstlisting}[language=cypher,basicstyle=\ttfamily\scriptsize]
MATCH (src1)-[:r]->(dst1),
     (src2)-[:r]->(dst2)
WHERE src1.dob1=src2.dob2 AND src1.company1=src2.company2 AND dst1.dob1=dst2.dob2 AND dst1.company1=dst2.company2 AND src1.graph='L' AND src2.graph='R' AND dst1.graph='L' AND dst2.graph='R'
CREATE p=(:U {id1:src1.UID, id2:src2.UID, MyGraphLabel:"U-"})-[:r]->(:U {id1:dst1.UID, id2:dst2.UID, MyGraphLabel:"U-"}) return p
UNION ALL
MATCH (src1), (src2)
OPTIONAL MATCH (src1)-[:r]->(dst1), (src2)-[:r]->(dst2)
WHERE src1.dob1=src2.dob2 AND src1.company1=src2.company2 AND src1.graph='L' AND src2.graph='R' AND dst1.graph='L' AND dst2.graph='R' AND ((dst1.dob1<>dst2.dob2 OR dst1.company1<>dst2.company2))
CREATE p=(:U {id1:src1.UID, id2:src2.UID,  MyGraphLabel:"U-"}) return p
UNION ALL
MATCH (src1)-[:r]->(dst1), (src2)
OPTIONAL MATCH (dst2)
WHERE src1.dob1=src2.dob2 AND src1.company1=src2.company2 AND dst1.dob1=dst2.dob2 AND dst1.company1=dst2.company2 AND src1.graph='L' AND src2.graph='R' AND dst1.graph='L' AND dst2.graph='R' AND ((NOT ((src2)-[:r]->())))
CREATE p=(:U {id1:src1.UID, id2:src2.UID,  MyGraphLabel:"U-"})-[:r]->(:U {id1:dst1.UID, id2:dst2.UID, MyGraphLabel:"U-"}) return p
UNION ALL
MATCH (src1), (src2)-[:r]->(dst2)
OPTIONAL MATCH (dst1)
WHERE src1.dob1=src2.dob2 AND src1.company1=src2.company2 AND dst1.dob1=dst2.dob2 AND dst1.company1=dst2.company2 AND src1.graph='L' AND src2.graph='R' AND dst1.graph='L' AND dst2.graph='R' AND ((NOT ((src1)-[:r]->())))
CREATE p=(:U {id1:src1.UID, id2:src2.UID,  MyGraphLabel:"U-"})-[:r]->(:U {id1:dst1.UID, id2:dst2.UID, MyGraphLabel:"U-"}) return p
UNION ALL
MATCH (src1)-[:r]->(dst1),
(src2)-[:r]->(dst3)
OPTIONAL MATCH (dst2)
WITH src1, dst1, src2, COLLECT(dst3) as coll, dst2
WHERE src1.graph='L' AND src2.graph='R' AND dst1.graph='L' AND dst2.graph='R' AND src1.dob1=src2.dob2 AND src1.company1=src2.company2 AND dst1.dob1=dst2.dob2 AND dst1.company1=dst2.company2 AND NONE (x IN coll WHERE dst1.dob1=x.dob2 AND dst1.company1=x.company2)
CREATE p=(:U {id1:src1.UID, id2:src2.UID,  MyGraphLabel:"U-"})-[:r]->(:U {id1:dst1.UID, id2:dst2.UID, MyGraphLabel:"U-"}) return p
UNION ALL
MATCH (src1)-[:r]->(dst3),
(src2)-[:r]->(dst2)
OPTIONAL MATCH (dst1)
WITH src1, dst1, src2, COLLECT(dst3) as coll, dst2
WHERE src1.graph='L' AND src2.graph='R' AND dst1.graph='L' AND dst2.graph='R' AND src1.dob1=src2.dob2 AND src1.company1=src2.company2 AND dst1.dob1=dst2.dob2 AND dst1.company1=dst2.company2 AND NONE (x IN coll WHERE x.dob1=dst2.dob2 AND x.company1=dst2.company2)
CREATE p=(:U {id1:src1.UID, id2:src2.UID,  MyGraphLabel:"U-"})-[:r]->(:U {id1:dst1.UID, id2:dst2.UID, MyGraphLabel:"U-"}) return p
\end{lstlisting}\caption{Cypher implementation for the graph disjunctive equi-join operator. Please note that, in addition to the limiations of the implementations of the conjunctive semantics, in the disjunctive we have an increase of patterns that need to be tested, created, and returned.}
			\label{fig:CypherEquiDisJoin}
		\end{minipage}

\end{figure}
	Please notice that $A_1$ and $A_2$ explicitly refer to the final graphs' attributes appearing only on one graph operand.
	Moreover, for each possible graph join operator, we have to specify which vertices are going to be
	linked in the final graph and whose nodes are going to have no neighbours. 
	
	Among all the possible graph query languages over property graph model, we consider Cypher. An example of the implementation of the an equi-join operator is provided in Figure \ref{fig:CypherEquiJoin}: the
	\texttt{CREATE} clause has to be used to generate new vertices and edges from graph patterns
	extracted through the \texttt{MATCH...WHERE} clause, and intermediate results are merged with
	\texttt{UNION ALL}.
	While current graph query languages allow to express our proposed graph
	join operator as a combination of the aforementioned operators,  our study shows that our specialized graph join
	algorithm outperforms the evaluation of the graph join with existing graph and relational query languages. We can also draw similar considerations for the disjunctive semantics where, in addition to the previous considerations, the number of the patterns that needs to be visited increases (Fig. \ref{fig:CypherEquiDisJoin}). 
	\medskip
	
	\begin{figure}[!t]
		\begin{minipage}[t]{\textwidth}
\begin{lstlisting}[language=sparql,basicstyle=\ttfamily\scriptsize]
CONSTRUCT { 
	?newSrc <url/to/graph> "Result";
	      <url/to/edges/result> ?newDst; 
	      <url/to/property/Ip1> ?ip1;
	      <url/to/property/Organization1> ?org1;
	      <url/to/property/Year1> ?y1;
	      <url/to/property/Ip2> ?ip2;
	      <url/to/property/Organization2> ?org2;
	      <url/to/property/Year2> ?y2.
	?newDst <url/to/graph> "Result";
	      <url/to/property/Ip1> ?ip3;
	      <url/to/property/Organization1> ?org3;
	      <url/to/property/Year1> ?y3;
	      <url/to/property/Ip2> ?ip4;
	      <url/to/property/Organization2> ?org4;
	      <url/to/property/Year2> ?y4.
} 
FROM NAMED <leftpath/to/graph>
FROM NAMED <rightpath/to/graph>
WHERE
{
  GRAPH ?g { 
  		?src1 <url/to/property/Id> ?id1;
	      <url/to/property/Ip1> ?ip1;
	      <url/to/property/Organization1> ?org1;
	      <url/to/property/Year1> ?y1.
  	}.
  GRAPH ?h { 
  		?src2 <url/to/property/Id> ?id2;
	      <url/to/property/Ip2> ?ip2;
	      <url/to/property/Organization2> ?org2;
	      <url/to/property/Year2> ?y2.
  	}
  filter(?g=<leftpath/to/graph> && 
         ?h=<rightpath/to/graph> &&
         ( ?org1 = ?org2 ) && ( ?y1 =  ?y2 ))
         
  BIND (URI(CONCAT("url/to/values/",?id1,"-",?id2)) AS ?newSrc)
  
  OPTIONAL {
  		GRAPH ?g { 
  			?src1 <url/to/edges/edge> ?dst1.
  			?dst1 <url/to/property/Id> ?id3;
	      <url/to/property/Ip1> ?ip3;
	      <url/to/property/Organization1> ?org3;
	      <url/to/property/Year1> ?y3.
  		}.
		GRAPH ?h { 
			?src2 <url/to/edges/edge> ?dst2.
			?dst2 <url/to/property/Id> ?id4;
	      <url/to/property/Ip2> ?ip4;
	      <url/to/property/Organization2> ?org4;
	      <url/to/property/Year1> ?y4.
  		}
		FILTER ( ( ?org3 = ?org4 ) && ( ?y3 = ?y4 ) )
		BIND (URI(CONCAT("url/to/values/",?id3,"-",?id4)) AS ?newDst)
	}
}
\end{lstlisting}
			\caption{SPARQL implementation for the graph conjunctive equi-join operator. Please note that while the SPARQl engine can optimize the graph traversal tasks, the creation of a new graph as a result is not included in the optimization steps.}
			\label{fig:SparqlEquiJoin}
		\end{minipage}
	\end{figure}
	On the other hand, in the case of RDF graph models, we have to discriminate whether vertices either represent entities or
	values that describe them and, consequently, we have to discriminate between edges representing
	relations among entities and the ones  acting as  attributes (when such each links an entity to its associated value
	expressed as the destination vertex.
	Even in this case the \textit{join condition} depends upon the specific graphs' schema,
	that may vary on different RDF graphs. In particular, as showed in Figure \ref{fig:SparqlEquiJoin}, SPARQL allows to access multiple graph resources
	through \textit{named graphs} and performs graph traversals one graph at a time through
	\textit{path joins}.
	At this point the \texttt{CONSTRUCT} clause is required if we
	want to finally combine the traversed paths from both graphs into a resulting graph.
	
	Consequently, for the two following conditions current graph query languages do not support
	graph joins as a primitive operator.
	\begin{enumerate}
		\item An explicit graph join operator is missing in current graph database languages.
		\item Even if we hold fixed the $\theta$ binary property, each time that the underlying graph schema
		changes we have to rewrite the join query every time, either because we have to specify how to merge
		the nodes and how to create final edges, or because we have to re-write the \textit{join condition}.
	\end{enumerate}

\subsection{PostgreSQL Scripts for Graph Joins and Incremental Updates}
\begin{figure}
\begin{lstlisting}[language=sql,basicstyle=\ttfamily\scriptsize]
-- Parsing the gender with a specific type
CREATE DOMAIN gender CHAR(1) CHECK (value IN ( 'F' , 'M' ) );
-- Starts the benchmark
\timing
-- Vertices for the left operand
create table lv(id serial not null, sex gender, name varchar, surname varchar, dob varchar, email varchar, company varchar, residence varchar);
-- Edges for the left operand
create table le(src integer, dst integer);
-- Vertices for the right operand
create table rv(id serial not null, sex gender, name varchar, surname varchar, dob varchar, email varchar, company varchar, residence varchar);
-- Edges for the right operand
create table re(src integer, dst integer);
-- Loading the table directly from the CSV format, using PostgreSQL's native methods
\copy lv(id,sex,name,surname,dob,email,company,residence) from '/path/to/dataset/krongen/diffs/diff_0_5__vertices.csv' DELIMITER ',' CSV HEADER;
\copy le(src,dst) from '/path/to/dataset/krongen/diffs/diff_0_5_' delimiter E'\t'  CSV;
\copy rv(id,sex,name,surname,dob,email,company,residence) from '/path/to/dataset/krongen/diffs/diff_0_6__vertices.csv' DELIMITER ',' CSV HEADER;
\copy re(src,dst) from '/path/to/dataset/krongen/diffs/diff_0_6_' delimiter E'\t'  CSV;
\end{lstlisting}
\caption{PostgreSQL's data loading for the graph join operands.}\label{SqlLoad}
\end{figure}
This section provides the main SQL commands in PostgreSQL's dialect that were exploited for benchmarking graph joins over relational databases. First, we show how we exploited customary PostgreSQL data loading operations from CSV files. As a consequence, no additional overhead was introduced while benchmarking (Figure \ref{SqlLoad}). The representation of vertices and edges for each single graph operand as distinct tables is the one suggested in \cite{sqlgraph}.

\begin{figure}[!t]
\begin{lstlisting}[language=sql,basicstyle=\ttfamily\scriptsize]
-- Vertex Querying (same for both Conjunctive and Disjunctive semantics)
explain analyse        
select lv.id as l, rv.id as r
from lv, rv
where lv.dob = rv.dob and lv.company = rv.company;

-- Edge Querying (Conjunctive)
create view vertexjoin as 
select lv.id as l, rv.id as r
from lv, rv
where lv.dob = rv.dob and lv.company = rv.company;

explain analyse
select src.l, src.r, dst.l, dst.r
from (select lv.id as l, rv.id as r
      from lv, rv
      where lv.dob = rv.dob and lv.company = rv.company) src,
     (select lv.id as l, rv.id as r
      from lv, rv
      where lv.dob = rv.dob and lv.company = rv.company) dst,
     le, re
where src.l = le.src and 
dst.l = le.dst and
src.r = re.src and
dst.r = re.dst;

-- Edge Querying (Disjunctive)
explain analyse
select src.*, dst.*
from le, re, vertexjoin as src, vertexjoin as dst
where (src.l = le.src and  dst.l = le.dst  and  src.r = re.src and dst.r = re.dst) 
UNION
select src.*, dst.*
from le, vertexjoin as src, vertexjoin as dst
WHERE  src.l = le.src and dst.l = le.dst and NOT EXISTS (select * from re where src.r = re.src and dst.r = re.dst)
UNION
select src.*, dst.*
from re, vertexjoin as src, vertexjoin as dst
WHERE  src.r = re.src and dst.r = re.dst and NOT EXISTS (select * from le where src.l = le.src and  dst.l = le.dst)
\end{lstlisting}
	\caption{PostgreSQL's data loading for the graph join operands.}\label{SqlJoin}
\end{figure}
Next, we show the queries exploited for the conjunctive and disjunctive algorithm (Figure \ref{SqlJoin}); please note that the commands \textit{explain analyse} were explicitly used for both analysing the  associated query plan, as well as avoiding to consider the results' query output from the total query time. Given that in the relational model vertices and edges can be represented as two separate tables, we separately benchmarked vertex and edge creation, while summing those two results up. Please observe that non-materialised view creations allows to both simplify the query code while exploiting explicit query rewriting operations.

\begin{figure}[!t]
	\begin{lstlisting}[language=sql,basicstyle=\ttfamily\scriptsize]
-- Vertex querying => creating the associated view
create materialized view vertexjoin as select lv.id as l, rv.id as r from lv, rv where lv.dob = rv.dob and lv.company = rv.company;
	
-- Edge querying
create materialized view edgejoin as                              
select src.l as src_l, src.r as src_r, dst.l as dst_l, dst.r as dst_r
from le, re, vertexjoin as src, vertexjoin as dst
where (src.l = le.src and  dst.l = le.dst  and  src.r = re.src and dst.r = re.dst);

-- Now, loading the incremental data!
\copy rv(id,sex,name,surname,dob,email,company,residence) from '/path/to/dataset/krongen/diffs/diff_1_6__vertices.csv' DELIMITER ',' CSV HEADER;
\copy re(src,dst) from '/path/to/dataset/krongen/diffs/diff_1_6_' delimiter E'\t'  CSV;
-- Actually performing the view refresh, without necessarily running the SQL query!
REFRESH MATERIALIZED VIEW vertexjoin;
REFRESH MATERIALIZED VIEW edgejoin;
\end{lstlisting}
	\caption{Running Graph Conjunctive EquiJoins with (positive) incremental updates in SQL.}\label{SqlIncr}
\end{figure}
Last, we provide the script in Figure \ref{SqlIncr} for running the incremental graph join updates for the conjunctive semantics, showing that data loads exploited customary PostgreSQL commands.

\bibliographystyle{abbrv}
\bibliography{bibliography.bib}

\begin{thebibliography}{1}

\bibitem{atzeniIT}
P.~Atzeni, S.~Ceri, S.~Paraboschi, and R.~Torlone.
\newblock {\em Database Systems - Concepts, Languages and Architectures (in
  {Italian})}.
\newblock McGraw-Hill, Milan, 3rd edition, 2009.

\bibitem{BergamiBM21}
G.~Bergami.
\newblock On efficiently equi-joining graphs.
\newblock In {\em {(to appear in) IDEAS'21}}. {ACM}.

\bibitem{rolleke94equivalencesof}
T.~R\"{o}lleke.
\newblock Equivalences of the probabilistic relational algebra.
\newblock Technical report, 1994.

\bibitem{sqlgraph}
W.~Sun, A.~Fokoue, K.~Srinivas, A.~Kementsietsidis, G.~Hu, and G.~Xie.
\newblock {SQLGraph}: An efficient relational-based property graph store.
\newblock In {\em Proceedings of the 2015 ACM SIGMOD International Conference
  on Management of Data}, pages 1887--1901.

\end{thebibliography}

\end{document}